\documentclass[aps, prl, twocolumn, amsmath, amssymb, superscriptaddress, reprint, longbibliography]{revtex4-2}

\usepackage[utf8]{inputenc}
\usepackage[T1]{fontenc}
\usepackage[colorlinks=true,citecolor=blue]{hyperref}
\usepackage{graphicx}
\usepackage{units}
\usepackage[dvipsnames]{xcolor}
\newcommand{\cbox}[2][]{{#2}}

\newcommand{\Tc}{T_\mathrm c}
\newcommand{\kB}{k_\mathrm B}
\newcommand{\angstrom}{\text{\AA}\vphantom{A}}

\graphicspath{ {./images/} }

\begin{document}


\title{Dimer Coupling Energies of the Si(001) Surface}


\author{Christian Brand}
\email[Corresponding author: ]{christian.brand@uni-due.de}
\affiliation{Faculty of Physics, University of Duisburg-Essen, 47057 Duisburg, Germany}

\author{Alfred Hucht}
\affiliation{Faculty of Physics, University of Duisburg-Essen, 47057 Duisburg, Germany}
\affiliation{Center for Nanointegration (CENIDE), University of Duisburg-Essen, 47057 Duisburg, Germany}

\author{Giriraj Jnawali}
\altaffiliation[Current address: ]{Department of Physics, University of Cincinnati, Cincinnati, OH 45221, USA}
\affiliation{Faculty of Physics, University of Duisburg-Essen, 47057 Duisburg, Germany}

\author{\firstname{Jonas D.} Fortmann}
\affiliation{Faculty of Physics, University of Duisburg-Essen, 47057 Duisburg, Germany}

\author{Bj\"{o}rn Sothmann}
\affiliation{Faculty of Physics, University of Duisburg-Essen, 47057 Duisburg, Germany}
\affiliation{Center for Nanointegration (CENIDE), University of Duisburg-Essen, 47057 Duisburg, Germany}

\author{Hamid Mehdipour}
\affiliation{Faculty of Physics, University of Duisburg-Essen, 47057 Duisburg, Germany}

\author{Peter Kratzer}
\affiliation{Faculty of Physics, University of Duisburg-Essen, 47057 Duisburg, Germany}
\affiliation{Center for Nanointegration (CENIDE), University of Duisburg-Essen, 47057 Duisburg, Germany}

\author{Ralf Sch\"{u}tzhold}
\affiliation{Institute of Theoretical Physics, Dresden University of Technology, 01062 Dresden, Germany}
\affiliation{Helmholtz-Zentrum Dresden-Rossendorf, 01328 Dresden, Germany}

\author{Michael \surname{Horn-von Hoegen}}
\affiliation{Faculty of Physics, University of Duisburg-Essen, 47057 Duisburg, Germany}
\affiliation{Center for Nanointegration (CENIDE), University of Duisburg-Essen, 47057 Duisburg, Germany}

\date{\today}


\begin{abstract}
The coupling energies between the buckled dimers of the Si(001) surface were determined through analysis of the anisotropic critical behavior of its order-disorder phase transition. 
Spot profiles in high-resolution low-energy electron diffraction as a function of temperature were analyzed within the framework of the anisotropic two-dimensional Ising model.
The validity of this approach is justified by the large ratio of correlation lengths, $\xi_\parallel^+/\xi_\perp^+ = 5.2$ of the fluctuating $c(4 {\times} 2)$ domains above the critical temperature $\Tc = \unit[(190.6 \pm 10)]{K}$.
We obtain effective couplings $J_\parallel = \unit[(-24.9 \pm 1.3)]{meV}$ along the dimer rows and $J_\perp = \unit[(-0.8 \pm 0.1)]{meV}$ across the dimer rows, i.e., antiferromagnetic-like coupling of the dimers with $c(4 {\times} 2)$ symmetry.
\end{abstract}


\maketitle




\par
\begin{figure*}[ht]
\centering
\includegraphics[scale=1.0]{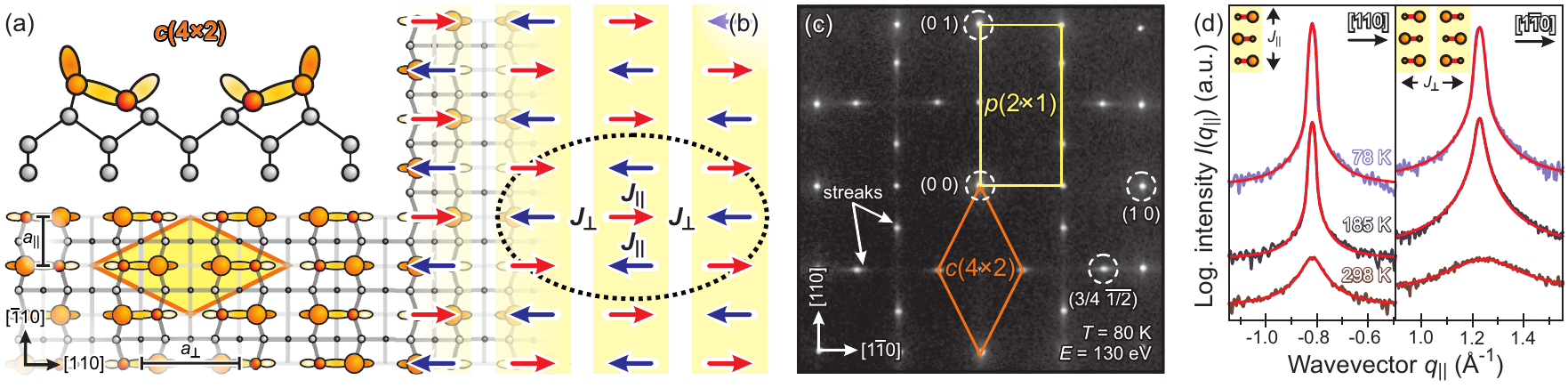}
\caption{
\textbf{Dimer reconstruction of Si(001).}
(a) Atomic structure model of the Si(001)-$c(4 {\times} 2)$ surface reconstruction.
(b) Spin model describing the arrangement of alternation of dimer buckling.
Coupling energies $J_\parallel$ along dimer rows and $J_\perp$ across the dimer rows are indicated.
(c) SPA-LEED pattern taken at $T = \unit[80]{K}$.
Primitive unit cells of the $c(4 {\times} 2)$ and $(2 {\times} 1)$ reconstruction are indicated by the orange rhombus and yellow rectangle, respectively.
(d) Intensity line profiles through the $(3/4~\overline{1/2})$ spot along (left, along $\left[ 1 1 0 \right]$) and across (right, along $\left[ 1 \overline{1} 0 \right]$) the dimer rows for selected temperatures below, around and above $\Tc$, respectively.
Experimental data (Fourier-filtered, 80\% of linear background subtracted) is plotted as purple to brownish lines, while corresponding fits are given by red lines.
Profiles are vertically shifted for better visibility.
}
\label{Fig.LEEDpatterns}
\end{figure*}

While the (001)-face of single crystalline silicon belongs to the most important surfaces both in technology and science, some of its fundamental properties are experimentally still unexplored.
The bare Si(001) surface exhibits a rich hierarchy of structural motives minimizing the surface free energy \cite{Ramstad:PRB51.14504, Alerhand:PRB35.5533, Zhu:PRB40.11868}.
Its structural key element are dimers composed of two Si surface atoms.
These dimers arrange in parallel dimer rows giving rise to a strong lattice and electronic anisotropy of the surface \cite{DabrowskiMuessig:SiSurfInt}.
Driven by a Jahn-Teller distortion the dimers become asymmetrically buckled, as sketched in Fig.~\ref{Fig.LEEDpatterns}(a) \cite{Ramstad:PRB51.14504, Chadi:PRL43.43}.
Since there are two choices for the buckling angle, a multitude of patterns, all based on $p(2 {\times} 1)$ as the smallest unit, may emerge.
At low temperatures, surface stress minimization causes alternating orientation of the dimer buckling angles along and across the dimer rows resulting in the $c(4 {\times} 2)$ reconstruction \cite{Dabrowski:PRB49.4790}.
The associated coupling energies between the dimers are subject of intense research, emphasized by a large number of theoretical calculations and being controversial at the limits of computational methods \cite{Ihm:PRL51.1872, Pillay:SurfSci554.150, Fu:SurfSci494.119, Xiao:PRM3.044410, Ramstad:PRB51.14504, Zhu:PRB40.11868, Alerhand:PRB35.5533, Khan:PRB39.3688, Low:PRB50.5352, Chadi:PRL43.43, Inoue:PRB49.14774, Gryko:PhysicaB194-196.381}.
These energies are dominated by short-ranged interactions between neighboring dimers and comparable to thermal energies.
Hence, experimentally accessible phase transitions are to be expected.
Accordingly, the continuous phase transition from the low-temperature ordered $c(4 {\times} 2)$ state (Fig.~\ref{Fig.LEEDpatterns}(a)) to the high-temperature disordered $p(2 {\times} 1)$ state is observed at a critical temperature $\Tc \approx \unit[200]{K}$ \cite{Tabata:SurfSci179.L63, Murata:PT53.125, Kubota:PRB49.4810, Matsumoto:PRL90.106103} and can be utilized for the experimental determination of the coupling energies.

\par
In this work, we employed spot-profile analysis low-energy electron diffraction (SPA-LEED) to follow the critical behavior of this order-disorder phase transition. 
The data analysis was done in two steps: the changes in spot profile of the $c(4 {\times} 2)$ spots were analyzed in the framework of the 
two-dimensional (2D) Ising universality class,
leading to accurate values for $\Tc=\unit[190.6]{K}$ as well as for the critical correlation length ratio $\xi_\parallel^+/\xi_\perp^+ = 5.2$. 
In a second step, these values were mapped onto the anisotropic 2D Ising model, from which
we obtained values for the effective coupling energies of $J_\parallel = \unit[-24.9]{meV}$ along and $J_\perp = \unit[-0.8]{meV}$ across the dimer rows with unprecedented precision.

\par
An appropriate description of the dimerized Si(001) surface structure is realized by mapping onto the aniso\-tropic 2D Ising model on a rectangular lattice.
The two states of the Ising spins $\sigma_{i,j} = \pm 1$ correspond to the two buckling orientations of the Si dimers as sketched in Fig.~\ref{Fig.LEEDpatterns}(b).
These two dimer configurations are separated by an energy barrier of $E_\mathrm{b} \approx \unit[90]{meV}$ \cite{Dabrowski:ASS56-58.15} which is large compared to the thermal energy near the critical point such that intermediate states are exponentially suppressed.
The Hamiltonian of the system is given by
\begin{equation}
    \mathcal{H} = - \sum_{i,j} \left( J_\parallel \sigma_{i,j} \sigma_{i,j+1} + J_\perp \sigma_{i,j} \sigma_{i+1,j} \right)\,,
\label{eq:IsingHamiltonian}
\end{equation}
where $J_\parallel$ is the effective exchange coupling between the nearest-neighbor dimers along the dimer row, while $J_\perp$ is the effective exchange coupling between neighboring rows.
The absence of next-nearest-neighbor couplings in \eqref{eq:IsingHamiltonian} is a consequence of the large correlation-length anisotropy of the considered system and will be justified below.
\par
The anisotropic 2D Ising model was solved analytically by Onsager \cite{Onsager:PR65.117}.
It exhibits a continuous phase transition at the critical temperature $\Tc$ determined by \cite{KramersWannier:PR60.252}
\begin{equation}
    \sinh \left( \frac{2|J_\parallel|}{\kB \Tc} \right) \, \sinh \left( \frac{2|J_\perp|}{\kB \Tc} \right) = 1\,.
\label{eq:OnsagerEquation}
\end{equation}
In the vicinity of the phase transition, the system exhibits universal critical behavior \cite{Kadanoff:Physics.2.263}, i.e., quantities such as the correlation length $\xi_\delta$ in direction $\delta \in \{\parallel,\perp\}$, the order parameter $\Psi$, and the susceptibility $\chi$ asymptotically behave as power laws of the reduced temperature $t = T/\Tc - 1$ as
\begin{subequations}
    \begin{align}
        \xi_\delta(t) &\simeq \xi_\delta^\pm \left| t \right|^{-\nu} , \label{eq:xi}\\ 
        \Psi(t) &\simeq \Psi^- \left( -t \right)^\beta , \label{eq:Psi}\\ 
        \chi(t) &\simeq \chi^\pm \left| t \right|^{-\gamma}\label{eq:chi}\,. 
    \end{align}
\label{eq:CritExp}%
\end{subequations}
Here, $\nu = 1$, $\beta = 1/8$ and $\gamma = 7/4$ are universal critical exponents within the 2D Ising universality class, \cbox{and} $\xi_\delta^\pm$, $\Psi^-$ and $\chi^\pm$ are the corresponding nonuniversal amplitudes above $(+)$ and below $(-)$ $\Tc$.

\par
The exact correlation lengths $\xi_\delta(T)$ of the anisotropic 2D Ising model above $\Tc$ in direction $\delta$ are given by \cite{McCoy+Wu:2D.Ising, HobrechtHucht:SciPostPhys7}
\begin{equation}
    \frac{\xi_\delta(T)}{a_\delta} \stackrel{T>\Tc}{=} \left[ \ln \coth \left( \frac{|J_\delta|}{\kB T} \right) - \frac{2 |J_{\bar\delta}|}{\kB T}\right]^{-1}\,,
\label{eq:CorrelationLength}
\end{equation}
where $\bar\delta$ denotes the direction perpendicular to $\delta$, while $a_\parallel = \unit[3.84]{\angstrom}$ and $a_\perp = 2 a_\parallel$ are the lattice parameters of the dimerized Si(001) surface. 
An expansion of Eqs.~\eqref{eq:CorrelationLength} around $\Tc$ from Eq.~\eqref{eq:OnsagerEquation} yields the correlation length amplitudes in Eq.~\eqref{eq:xi},
\begin{equation}\label{eq:CorrelationLengthAmplitudes}
    \frac{\xi^+_\delta}{a_\delta} = \left[\frac{2|J_{\delta}|}{\kB\Tc} \sinh\left(\frac{2|J_{\bar\delta}|}{\kB\Tc}\right) + \frac{2|J_{\bar\delta}|}{\kB\Tc}\right]^{-1}\,,
\end{equation}
from which one can deduce a simple relation between the coupling energies $J_\delta$ and the ratio of correlation length amplitudes,
\begin{equation}
        \sinh\left( \frac{2|J_\delta|}{\kB \Tc} \right) = \frac{\xi^+_\delta/a_\delta}{\xi^+_{\bar\delta}/a_{\bar\delta}}\,,
\label{eq:Couplings_from_ratio}
\end{equation}
such that we can determine the anisotropic coupling energies solely from the correlation length amplitude ratio \cite{LaBella:PRL84.4152}.
Note that the sign of $J_\delta$ has to be determined from the diffraction analysis below.



\par
Experimentally, we followed the order-disorder phase transition by means of spot profile analysis low-energy electron diffraction (SPA-LEED) which combines high resolution in reciprocal space with superior signal-to-noise ratio \cite{Scheithauer:SurfSci178.441, HvH:ZfK214.591}. 
The experiments were performed at ultra-high vacuum (UHV) conditions at a base pressure $p < \unit[2 \times 10^{-10}]{mbar}$ in order to ensure very low surface contamination through adsorption from residual gas.
The Si(001) sample (miscut $< 0.1^\circ$
{, Wacker Chemie AG, Burghausen}) was mounted on a cryostat for sample cooling by liquid nitrogen.
Direct current was applied to heat the sample for degassing at $600^\circ$C and subsequent flash-annealing at $T > 1200^\circ$C for \unit[5]{s} with the pressure remaining in the $\unit[10^{-10}]{mbar}$ regime.
Subsequently, the sample was rapidly cooled to \unit[78]{K}.
Using the built-in resistive heater of the cryostat, the sample was heated from \unit[78]{K} to \unit[400]{K} at a rate of \unit[10]{K/min}, while the sample temperature was measured using a Pt100 Ohmic sensor.
The systematic error in temperature determination is of the order of $\pm \unit[10]{K}$ while the statistical error is less than $\pm \unit[1]{K}$.
At the same time spot profiles through the (00) spot, four $p(2 {\times} 1)$ spots and one of the $c(4 {\times} 2)$ spots were continuously taken by SPA-LEED at an electron energy of $E = \unit[112]{eV}$.
The instrumental resolution of $\unit[(17.5 \pm 0.3) \times 10^{-3}]{\angstrom^{-1}}$ was determined from the sharpest spot of the pattern.
From the FWHM of the (00) spot, we estimated a mean terrace width larger than \unit[50]{nm} which is consistent with the expected terrace width of $\gtrsim \unit[150]{nm}$.
We also confirmed that during our SPA-LEED measurements at low beam current no disorder of the $c(4 {\times} 2)$ was induced by the electron beam as reported by others \cite{Shirasawa:PRL94.195502, Mizuno:PRB69.241306, Seino:PRL93.036101, Schmidt:CAP6.331}.
While in low-temperature scanning tunneling microscopy (STM) studies bias voltages above around \unit[0.2-1.4]{V} can induce phasons through flipping of dimers, leading to local $p(2 {\times} 2)$ structures \cite{Pennec:PRL96.026102, Sagisaka:PRB71.245319}, this effect can be excluded in our experiment since the beam current density was only $\unit[10-100]{nA/mm^2}$, i.e., many orders of magnitude lower than in typical STM experiments.



\begin{figure*}[ht]
\centering
\includegraphics[scale=1.0]{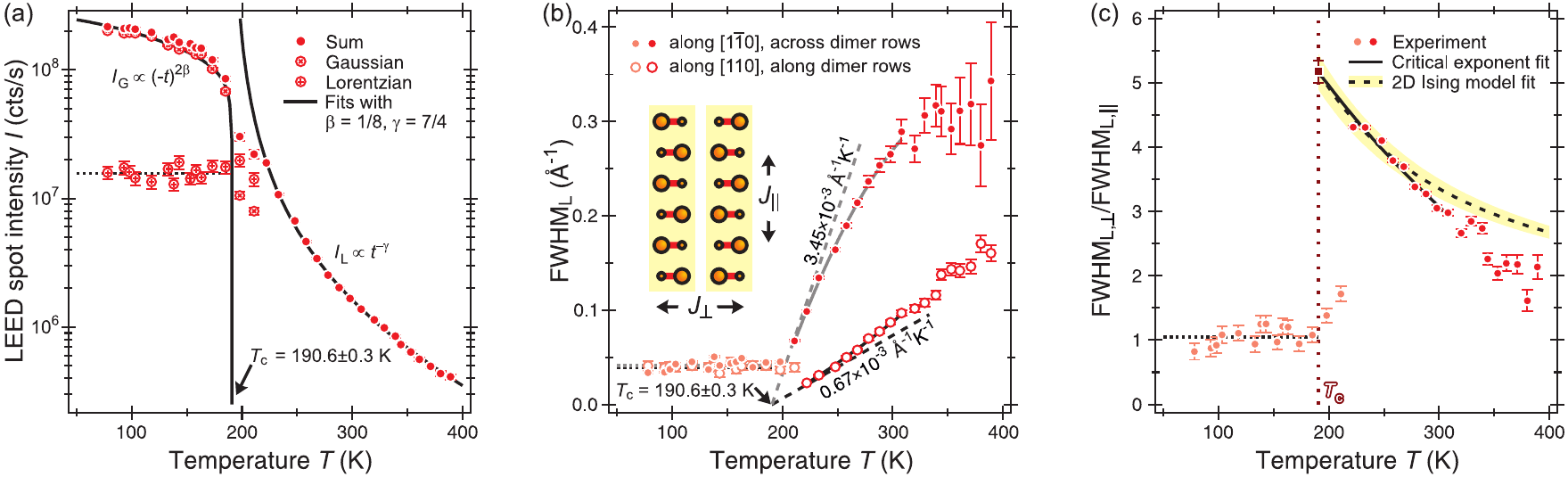}
\caption{
\textbf{Profile analysis of the $(3/4~\overline{1/2})$ spot.}
(a) Critical behavior of the Gaussian and Lorentzian contributions to the intensity during and above the phase transition, respectively.
Data are not corrected for the Debye-Waller effect ($\Theta_\mathrm{D} = \unit[(391 \pm 7)]{K}$ from fit to integrated intensity [data not shown]).
(b) Temperature-dependent Lorentzian FWHMs along (along $\left[ 1 1 0 \right]$) and across (along $\left[ 1 \overline{1} 0 \right]$) the dimer rows, respectively.
(c) Ratio $\mathrm{FWHM}_{\mathrm{L},\perp} / \mathrm{FWHM}_{\mathrm{L},\parallel}$ of the Lorentzian contribution.
Solid lines in (a-c) indicate fits to the critical behavior predicted by Onsager theory, determining $\Tc = \unit[(190.6 \pm 0.3)]{K}$.
First order corrections are taken into account for the FWHMs, respectively.
Their asymptotic (linear) behavior is shown by dashed lines in (b).
The expected behavior for the anisotropic 2D Ising model (dashed line in (c)) is derived by matching values with the critical behavior at $\Tc$.
The yellow area accounts for the systematic error of the measurement for the Ising model fit, while the brown square at $\Tc$ marks the crossing point including statistical errors.
Pink data points and dotted lines (fits) belong to the so-called domain state and are not taken into account for the fits of the critical behavior.
}
\label{Fig.Int&FWHM}
\end{figure*}

\par
Figure~\ref{Fig.LEEDpatterns}(c) shows a LEED pattern of the surface taken with $E = \unit[130]{eV}$ at \unit[80]{K}, i.e., below the phase transition temperature.
It exhibits sharp diffraction spots and low background reflecting the low step density and low defect and adsorbate density.
Intensity line profiles through the (00) spot and half-integer order spots exhibit sharp Gaussian-shaped spots reflecting the long-range order of the surface.
These spots exhibit no temperature dependence besides a Debye-Waller behavior.
The pattern is composed of an incoherent superposition of two distinct $c(4 {\times} 2)$ patterns, originating from the by $90^\circ$-rotated dimer rows on adjacent terraces.
While the fourfold periodicity in the diffraction pattern refers to the direction across the dimer rows, the $\times 2$ periodicity is along the dimer rows.
The streak-like intensity centered at the quarter-integer order spot positions is even visible far above the phase transition temperature in LEED \cite{Kubota:PRB49.4810, Murata:PT53.125} and He ion scattering \cite{Cardillo:PRL40.1148, Cardillo:PRB21.1497} and is indicative for fluctuations of the dimers, i.e., activation of diffusive phase defects (so-called phasons) \cite{Pennec:PRL96.026102, Kawai:JPSJ68.3936, Natori:ASS212-213.705, Hafke:PRL124.016102}.

\par
Using intensity line profiles through the $(3/4~\overline{1/2})$ spot which is a measure for the alternating order of the dimers, i.e., the antiferromagnetic order in the anisotropic 2D Ising model, both along ($\left[ 1 1 0 \right]$ direction) and across ($\left[ 1 \overline{1} 0 \right]$ direction) the Si dimer rows, the temperature dependence of the spot profile was recorded.
Exemplarily, six line profiles (purple to brownish lines) and their respective fits (red lines) are shown in Fig.~\ref{Fig.LEEDpatterns}(d).
The spot intensity $I(\mathbf{q},t)$ exhibits a sharp drop at $\approx \unit[200]{K}$ indicative for the phase transition (see Fig.~\ref{Fig.Int&FWHM}(a)) \cite{Tabata:SurfSci179.L63, Murata:PT53.125}.
In accordance with the 2D Ising model and Refs.~\cite{Kubota:PRB49.4810, Murata:PT53.125}, the line profile with the spot at reciprocal lattice vector $\mathbf{q}_0$ was fitted by the sum of a peak
\begin{equation}
    I(\mathbf{q},t) = A^-_\mathrm{G} \delta(\mathbf{q}-\mathbf{q}_0) \left( -t \right)^{2\beta} + A^\pm_\mathrm{L}(\mathbf{q}-\mathbf{q}_0) \left| t \right|^{-\gamma}
\label{eq:SpotProfile}
\end{equation}
with amplitudes $A^\pm_\mathrm{G,L}$, and a linear background.
Here, $I^-_\mathrm{G} = A^-_\mathrm{G} \left( -t \right)^{2\beta}$ is the sharp central $\delta$-spike (Gaussian-shaped contribution with $\mathrm{FWHM}_{\mathrm{G},\delta}$) proportional to the square of the order parameter $\Psi(t)$ from Eq.~\eqref{eq:Psi} and following a power law with exponent $2\beta = 1/4$.
Accordingly, $I^\pm_\mathrm{L} = A^\pm_\mathrm{L} \left| t \right|^{-\gamma}$ is the broad diffuse part (Lorentz\-ian-shaped contribution with $\mathrm{FWHM}_{\mathrm{L},\delta} = 2\pi/\xi_\delta$) of the spot profile above and below $\Tc$, which is proportional to the susceptibility $\chi(t)$ from Eq.~\eqref{eq:chi} and scales with an exponent of $\gamma = 7/4$.
To account for the instrumental response function of the SPA-LEED, a pseudo-Voigtian function (sum of a Lorentzian and a Gaussian peak with the minimum Gaussian FWHM of the sharpest spot) was used to fit the Lorentzian contribution.

\par
Below $T \approx \unit[200]{K}$ the spot profile of the $(3/4~\overline{1/2})$ spot consists of a sharp Gaussian and a weaker constant Lorentzian contribution, depicted by $\otimes$ and $\oplus$ in Fig.~\ref{Fig.Int&FWHM}(a), respectively.
The FWHMs of both contributions are small, isotropic and constant below $\Tc$.
Above $T \approx \unit[200]{K}$ the central Gaussian spike has disappeared, while the width of the broad diffuse part strongly increases, but is still clearly visible at room temperature and above.
The intensities of the line profiles of the $(3/4~\overline{1/2})$ spot were corrected for the Debye-Waller effect with $\Theta_\mathrm{D} = \unit[(391 \pm 7)]{K}$ as obtained from fits to integrated intensity of the spot.
These line profiles, namely the intensities of the Gaussian central spike $I_\mathrm{G}$ and of the broad diffuse Lorentzian $I_\mathrm{L}$, as well as the Lorentzian peak widths $\mathrm{FWHM}_{\mathrm{L},\delta}$ are further analyzed and compared to the theoretical predictions of the anisotropic 2D Ising model.
Both intensity contributions vary strongly as functions of temperature, i.e., reflecting the critical behavior of the phase transition.
We derived $\Tc = \unit[(190.6 \pm 0.3)]{K}$ (statistical error only) by a global fit for all four critical quantities (solid lines in Fig.~\ref{Fig.Int&FWHM}), i.e., $I_\mathrm{G,L}$ and $\mathrm{FWHM}_{\mathrm{L},\delta}$.

\par
The Lorentzian FWHMs along and across the dimer rows are shown in Fig.~\ref{Fig.Int&FWHM}(b).
For $T \lesssim \unit[200]{K}$ (pink data points), we observe a quenched domain state, i.e., finite-sized $c(4 {\times} 2)$ domains with constant $\mathrm{FWHM}_{\mathrm{L},\delta} = \unit[(40 \pm 1) \times 10^{-3}]{\angstrom^{-1}}$ (fit with dotted lines) which we attribute to nonequilibrium dynamics:
during the preparation of the $c(4 {\times} 2)$-reconstructed surface, the cooling rate was too fast for reaching the long-range-ordered state while passing the critical point.
The fluctuating dimer system is quenched into a nonequilibrium state.

\par
Above the critical temperature both Lorentzian FWHMs increase from zero asymptotically (dashed lines) with slopes $b_\parallel/\Tc = \unit[(0.67 \pm 0.04) \times 10^{-3}]{\angstrom^{-1} K^{-1}}$ and $b_\perp/\Tc = \unit[(3.45 \pm 0.20) \times 10^{-3}]{\angstrom^{-1} K^{-1}}$ (statistical plus systematic error).
The fit (solid lines in Fig.~\ref{Fig.Int&FWHM}(b)) to the data exhibits a clear deviation from linear behavior due to corrections to scaling, which are expected to be linear for the 2D Ising model \cite{HobrechtHucht:SciPostPhys7}, and are well described up to \unit[310]{K} as $\mathrm{FWHM}_{\mathrm{L},\delta}(t) = b_\delta t (1 + c_\delta t + \ldots)$.

\par
Even close to $\Tc$ the maximum observed correlation length was $\approx \unit[30]{nm}$.
Since this value is much smaller than the average terrace width of $\gtrsim \unit[150]{nm}$ the associated finite-size effects can be neglected here.

\par
Eventually, we obtain the coupling ratio $J_\parallel / J_\perp = 31.4 \pm 1.7$ by comparing the exact relation Eq.~\eqref{eq:Couplings_from_ratio} with the extrapolation $t \to 0^+$ of the experimentally observed temperature dependence of the Lorentzian FWHM ratio shown in Fig.~\ref{Fig.Int&FWHM}(c),
\begin{align}
\lim_{t\to 0^+}\frac{\mathrm{FWHM}_{\mathrm{L},\perp}(t)} {\mathrm{FWHM}_{\mathrm{L},\parallel}(t)} = \frac{b_\perp}{b_\parallel} = \frac{\xi^+_\parallel}{\xi^+_\perp} = 5.2 \pm 0.2\,.
\end{align}
With the estimated critical temperature $\Tc = \unit[190.6]{K}$, we finally find
\begin{gather*}
J_\parallel = \unit[(-24.9 \pm 1.3)]{meV} \; \mathrm{and} \; J_\perp = \unit[(-0.8 \pm 0.1)]{meV}\,,
\end{gather*}
where the error bars are due to systematic errors in temperature and FWHM ratio while the statistical errors are negligible.
The negative sign of both couplings follows from the spot positions in the diffraction pattern, leading to antiferromagnetic-like coupling of the dimers along and across the dimer rows with $c(4 {\times} 2)$ symmetry.
The critical temperature predicted by the more recent density functional theory calculations is in good agreement with the observed $\Tc$ \cite{Pillay:SurfSci554.150, Xiao:PRM3.044410}.
Theory and experiment agree that the intra-row interaction is much stronger than the inter-row interaction.
However, the ratio $J_\parallel / J_\perp$ of these interactions is very sensitive to the density functional used.
Thus, our precise experimental determination of this ratio provides a benchmark for future theoretical works.

\par
Finally, we comment on the validity of the considered nearest-neighbor (nn) Ising model: 
due to the huge correlation length anisotropy $(\xi_\parallel^+/a_\parallel)/(\xi_\perp^+/a_\perp) = 10.3$, Eq.~\eqref{eq:Couplings_from_ratio}, and the resulting pronounced short-range order in \emph{parallel} direction even near criticality, possible diagonal next-nearest-neighbor (nnn) couplings ($J_\mathrm{D} \sigma_{i,j} \sigma_{i \pm 1,j \pm 1}$) as well as nnn couplings in parallel direction ($J_\parallel^{(2)} \sigma_{i,j} \sigma_{i,j \pm 2}$) can safely be absorbed additively into renormalized effective nn couplings according to $J_\perp - 2 J_\mathrm{D} \mapsto J_\perp$ and $J_\parallel - J_\parallel^{(2)} \mapsto J_\parallel$, respectively, justifying the utilized nn Ising model \textit{a posteriori}.
We note that this additive approach is asymptotically correct for large correlation length anisotropy while for isotropic systems a more elaborate treatment is necessary \cite{Zandvliet:PhaseTrans:2009}.
These nnn couplings would only become relevant at much higher temperatures $T \gtrsim 2J_\parallel/\kB \approx \unit[580]{K}$.



\par
In conclusion, we use a two steps analysis to gain access to system parameters which are experimentally challenging to determine otherwise. 
In a first step, we use the known universal critical exponents of the 2D Ising universality class to accurately determine the critical temperature $\Tc$ and the correlation length ratio $\xi^+_\parallel/\xi^+_\perp$ of the system, see Fig.~\ref{Fig.Int&FWHM}.
In a second step, we map the results onto the exactly solvable anisotropic 2D Ising model to extract the effective coupling energies in the two directions.

\par
In detail, we employed the continuous order-disorder phase transition of Si(001) from the $c(4 {\times} 2)$ low temperature state to the $p(2 {\times} 1)$ high temperature state to determine the effective coupling energies between the alternately buckled Si dimers.
The clean and defect-free Si(001) surface exhibits critical behavior of intensity and correlation lengths obtained by means of high-resolution LEED with a critical temperature $\Tc = \unit[190.6]{K}$ and is evaluated in the framework of the anisotropic 2D Ising model.
From the ratio of the widths of diffuse intensity along and across the dimer rows we determined the effective coupling energies $J_\parallel = \unit[-24.9]{meV}$ and $J_\perp = \unit[-0.8]{meV}$.

\par
This work not only provides an answer to the long-standing question of the coupling energies of one of the world's most important surfaces, but also paves the road for application to other systems exhibiting phase transitions of the 2D Ising universality class such as the dimerized surfaces of Ge(001) \cite{Zandvliet:PhysRep388.1, Kevan:PRB32.2344, Lucas:PRB47.10375, Cvetko:SurfSci447.L147}, GaAs(001) \cite{LaBella:PRL84.4152}, and $\beta$-SiC(001) \cite{Aristov:PRL79.3700}, noble metal surfaces such as Au(110) \cite{Campuzano:PRL54.2684} or Pt(110) \cite{Zuo:JoVSTA8.2474}, Fe$_3$O$_4$(001) \cite{Bartelt:PRB88.235436}, adsorbate systems like O/W(112) \cite{Wang:PRB31.5918}, or even to other universality classes like for one-dimensional atomic wires such as Au/Si(553) \cite{Hafke:PRL124.016102}.


\section*{Author Contributions}

\par
G.J. performed the experiments.
C.B. analyzed the data and prepared the figures.
All authors discussed the results and drafted the manuscript.
M.H.-v.H., A.H., B.S., P.K., and R.S. conceived and supervised the project.
All authors have given approval to the final version of the manuscript.

\par
The authors declare no competing financial interest.


\section*{Acknowledgements}

\par
Funded by the Deutsche Forschungsgemeinschaft (DFG, German Research Foundation) through projects A02, B02, B03, B07, and C03 of Collaborative Research Center SFB1242 ``Nonequilibrium dynamics of condensed matter in the time domain'' (Project-ID 278162697).
We kindly acknowledge supporting LEEM measurements by F.-J. Meyer zu Heringdorf and D. Wall.


\bibliography{Bibliography.bib}



\end{document}